\documentstyle[epsfig]{elsart}
\begin{document}
\begin{frontmatter}
\hfill 
MZ-TH/01-20

\vspace{1cm}

\title{Semileptonic decays of double heavy baryons}

\author[tbn]{Amand  Faessler,}
\author[tbn]{Th. Gutsche,}
\author[jinr]{M.A. Ivanov,}
\author[mainz]{J.G. K\"{o}rner,}
\author[tbn]{V.E. Lyubovitskij}
\address[tbn]{Institut f\"{u}r Theoretische Physik,
Universit\"{a}t T\"{u}bingen, Auf der Morgenstelle 14,
D-72076 T\"{u}bingen, Germany}  
\address[jinr]{Bogoliubov Laboratory of Theoretical Physics, \\
Joint Institute for Nuclear Research, 141980 Dubna, Russia}
\address[mainz]{Institut f\"{u}r Physik, Johannes Gutenberg-Universit\"{a}t, 
D-55099 Mainz, Germany}

\begin{abstract}
We study the semileptonic decays of the lowest lying double heavy baryons 
using the relativistic three-quark model. We do not employ a heavy quark
mass expansion but keep the masses of the heavy quarks and baryons finite.
We calculate all relevant form factors and decay rates. 
\end{abstract}
       
\end{frontmatter}

The  semileptonic decays of heavy mesons and baryons are ideally suited 
to extract the Cabibbo-Kobayashi-Maskawa (CKM) matrix elements.
The heaviest flavored bottom-charm $B_c$-meson was observed by the 
CDF Collaboration \cite{CDF} in the analysis of the decay mode 
$B_c\to J/\psi \bar l \nu$. The discovery of the $B_c$-meson raises hopes 
that double heavy flavored baryons will also be discovered in the near future.
The theoretical treatment of the systems with two heavy quarks is complicated 
by the fact that one cannot make use of an expansion in terms of the inverse 
heavy quark masses. Previously, nonrelativistic potential models, diquark 
approximation, QCD sum rules and nonrelativistic QCD have been used to 
describe the  spectroscopy of double heavy baryons and to estimate the 
inclusive and some exclusive decay modes of such systems (for review, 
see \cite{Likh}-\cite{GMS} and references therein).

In  \cite{Bc} we have studied the semileptonic decays of the double heavy 
$B_c$-meson within a relativistic constituent quark model. The relativistic 
consti\-tu\-ent qu\-ark  model \cite{model}  can be viewed as an effective 
quantum field theory approach based on an  interaction Lagrangian of hadrons  
interacting with their constituent quarks. Universal and reliable predictions 
for exclusive processes involving both  mesons composed from a quark and 
antiquark and baryons composed from three quarks result from this approach. 
The coupling strength of hadrons $H$  to their constituent quarks
is determined by the compositeness condition $Z_H=0$ \cite{SWH,EI} where 
$Z_H$ is the wave function renormalization constant of the hadron. The 
quantity $Z_H^{1/2}$ is the matrix element between a physical particle state 
and the corresponding bare state. The compositeness condition $Z_H=0$ enables 
us to represent a bound state by introducing a hadronic field interacting 
with its constituents so that the renormalization factor is equal to zero. 
This does not mean that we can solve the QCD bound state equations but we are 
able to show that the condition $Z_H=0$ provides an effective and 
self-consistent way to describe the coupling of the particle to its 
constituents. One  starts with an effective interaction Lagrangian written 
down in terms of quark and hadron variables. Then, by using Feynman rules, 
the $S$-matrix elements describing hadron-hadron interactions are given in 
terms of a set of quark diagrams. In particular, the compositeness condition 
enables one to avoid the double counting of quark and hadron degrees of 
freedom. This approach is self-consistent and all calculations of physical
observables are straightforward. There is  a small set of  model parameters: 
the values of the constituent quark masses and the scale parameters that 
define the size of the distribution of the constituent quarks inside a given 
hadron. The shapes of the vertex functions and the  quark  propagators can in 
principle  be determined from an analysis of the Bethe-Salpeter (Fadde'ev) 
and Dyson-Schwinger equations, respectively, as done e.g. in \cite{DSE,DSEH}. 
In the present paper we, however, choose a more phenomenological approach 
where the vertex functions are modelled by Gaussian forms and the quark 
propagators are given by local representations. We have demonstrated in our 
papers \cite{model,RTQM} that the relativistic constituent model is consistent 
with the heavy quark symmetry in the limit of infinite quark masses.
We mention that the authors of \cite{Gatto} have developed a relativistic 
quark model approach for meson transitions which shows some similarities to 
our approach. They also use an effective heavy meson Lagrangian to describe 
the couplings of mesons to quarks. They use, however, point-like meson-quark 
interactions. Loop momenta are explicitly cut off at around 1 GeV in their 
approach \cite{Gatto}. In our approach we use momentum dependent meson-quark 
interactions which  provide for an effective cut-off of the loop integration.

We have elaborated the so-called {\it relativistic three-quark model} (RTQM)
to study the properties of heavy baryons containing a single heavy quark
(bottom or charm). For the heavy quarks we have used propagators appropriate
for the heavy quark limit. Various  observables describing semileptonic and 
nonleptonic decays as well as  one-pion and one-photon transitions have been 
successfully described  in this approach \cite{RTQM}. Recently, the RTQM was 
extended to include the effects of finite quark masses \cite{finite}.

In this paper we employ the RTQM \cite{finite} to calculate the form factors 
and widths of the semileptonic decays  of the lowest lying $\Xi_{bc}$ and 
$\Xi_{cc}$ baryons. We follow the strategy adopted in Ref.~\cite{Bc}
where we have studied leptonic and semileptonic decays of the $B_c$-meson. 
We employ the impulse approximation in calculating the matrix elements which 
previously has been widely used in phenomenological Dyson-Schwinger equation 
studies (see, e.g. Ref.~\cite{DSEH}). In the impulse approximation one 
assumes that the vertex functions depend only on the loop momentum flowing 
through the vertex.

We start with a brief description of our approach. As was mentioned above,
baryons are described as bound states of constituent quarks in the RTQM. 
The general form of the SU(5)-invariant lagrangian describing the interaction 
of three low-lying SU(5)-multiplets with their three-quark currents are 
written as
\begin{equation}\label{L_int}
{\cal L}_{\rm int}(x)=
{\cal L}_{\rm int}^{1/2^-}(x)+
{\cal L}_{\rm int}^{1/2^+}(x)+
{\cal L}_{\rm int}^{3/2^+}(x)
\end{equation}
where
\begin{eqnarray*}
{\cal L}_{\rm int}^{1/2^-}(x) &=& g_{F}
\bar F^{[m_1m_2m_3]}(x)\, J^{m_1m_2m_3}_{F}(x)\,+\,{\rm h.c.}\,,
\\
{\cal L}_{\rm int}^{1/2^+}(x) &=& g_{B}
\bar B^{[m_1m_2]m_3}(x)\, J^{m_1m_2m_3}_{B}(x)\,+\,{\rm h.c.}\,,
\\
{\cal L}_{\rm int}^{3/2^+}(x) &=& g_{D}
\bar D^{\{m_1m_2m_3\};\mu}(x)\, J^{m_1m_2m_3;\mu}_{D}(x)\,+\,{\rm h.c.}\,.
\end{eqnarray*}
Here $m_i=u,d,c,s,b$ are flavor indeces. According to the SU(5)-classification 
\begin{eqnarray*}
5 \otimes 5 \otimes 5  = 10_A \oplus 40_M \oplus 40_M \oplus 35_S 
\end{eqnarray*}
there is the  antisymmetric decuplet $F^{[m_1m_2m_3]}$ with  
$J^P={\frac{1}{2}}^-$, two 40-plets $B^{[m_2m_1]m_3}$ with mixed symmetry 
and $J^P={\frac{1}{2}}^+$ and a symmetric 35-plet $D^{\{m_1m_2m_3\}}$ with 
$J^P={\frac{3}{2}}^+$. 

The three-quark currents are written as
\begin{eqnarray*}
J^{m_1m_2m_3}_{F}(x)&=&\int\!dx_1\!\int\!dx_2\!\int\!dx_3\,
\Phi_F(x;x_1,x_2,x_3)\, \\
&&\times\gamma^\mu\gamma^5 q_{a_1}^{m_1}(x_1) 
\left(q_{a_2}^{m_2}(x_2)\,C\gamma^\mu\gamma^5\, q_{a_3}^{m_3}(x_3)\right) 
\varepsilon^{a_1a_2a_3}\,, \\
J^{m_1m_2m_3}_{B}(x)&=&\int\!dx_1\!\int\!dx_2\!\int\!dx_3\,
\Phi_B(x;x_1,x_2,x_3)\, \\
&&\times\gamma^\mu\gamma^5 q_{a_1}^{m_1}(x_1) 
\left(q_{a_2}^{m_2}(x_2)\,C\gamma^\mu\, q_{a_3}^{m_3}(x_3)\right) 
\varepsilon^{a_1a_2a_3} \,, \\
J^{m_1m_2m_3;\mu}_{D}(x)&=&\int\!dx_1\!\int\!dx_2\!\int\!dx_3\,
\Phi_D(x;x_1,x_2,x_3)\, \\
&&\times q_{a_1}^{m_1}(x_1) \left(q_{a_2}^{m_2}(x_2)\,C\gamma^\mu\, 
q_{a_3}^{m_3}(x_3)\right) \varepsilon^{a_1a_2a_3}\ 
\end{eqnarray*}
where $a_i=1,2,3$ are color indices. $F$, $B$ and $D$ denote the above three 
multiplets. Note that the function $\Phi_H$ is taken to be invariant under 
the translation $x\to x+a$ which guarantees Lorentz invariance for the 
interaction Lagrangian Eq.~(\ref{L_int}). SU(5)-symmetry is broken by 
employing explicit baryon and quark mass values when calculating matrix 
elements. In this paper we limit our attention to the basic semileptonic 
decay modes of the lowest lying $1/2^+$ double heavy baryons:
$\Xi_{\rm bcq}\to \Xi_{\rm bsq}\,+\,\bar l\nu$,
$\Xi_{\rm bcq}\to \Xi_{\rm ccq}\,+\,\bar l\nu$
and
$\Xi_{\rm ccq}\to \Xi_{\rm csq}\,+\,\bar l\nu$ (q=u or d).
The appropriate interpolating currents with diquarks in the symmetric state
and masses are shown in Table 1. The values of the masses are taken from 
potential models (see, for example, \cite{Likh,Faust}).

\begin{table}[t]
\caption{The lowest-lying states of  double heavy $\Xi$-baryons 
with diquarks in the symmetric state. The light quark $q$ denotes $u$ or $d$.}
\def\arraystretch{2}
\begin{center}
\begin{tabular}{|c|c|c|}
\hline
Baryon &Interpolating current & Mass (GeV) \\
\hline\hline
$\Xi_{cs}$    & 
-$\sqrt{2}\,\gamma^\mu\gamma^5 c_a\,(s_b C\gamma^\mu q_c)\varepsilon^{abc}$ & 
2.47 \\
\hline
$\Xi_{cc}$    & 
$\gamma^\mu\gamma^5 q_a\,(c_b C\gamma^\mu c_c)\varepsilon^{abc}$ & 3.61 \\
\hline
$\Xi_{bs}$    & 
-$\sqrt{2}\,\gamma^\mu\gamma^5 b_a\,(s_b C\gamma^\mu q_c)\varepsilon^{abc}$ & 
5.80 \\
\hline
$\Xi_{bc}$    & 
-$\sqrt{2}\,\gamma^\mu\gamma^5 b_a\,(c_b C\gamma^\mu q_c)\varepsilon^{abc}$ & 
7.00 \\
\hline\hline
\end{tabular}
\end{center}
\end{table}

The semileptonic transition amplitude is defined as
\begin{eqnarray}\label{ampl}
A\left(\Xi_{i}(p)\to\Xi_{f}(p')\,\bar l\nu\right)&=&
V_{if}\,\frac{G_F}{\sqrt{2}}\,
(\bar l O^\mu \nu)\,
\left(\bar u_{f}(p')\,
\Lambda^\mu_{i\to f}(p,p')\, u_{i}(p)\right) \,,
\end{eqnarray}
where $O^\mu=\gamma^\mu(1-\gamma^5)$. $V_{if}$ is the relevant element of the 
$CKM$-matrix where we use $V_{bc}=0.04$ and $V_{cs}=0.97$. The amplitude 
$\Lambda^\mu$ is decomposed into a set of six invariant  form factors which 
are functions of the momentum transfer squared $q^2=(p-p')^2$ only:
\begin{eqnarray}\label{def}
&&\Lambda^\mu_{i \to f}(p,p')= \\
&&\gamma^\mu\,(F_1^V-F_1^A\,\gamma^5)\,+\, i\sigma^{\mu\nu}q^\nu\, 
(F_2^V-F_2^A\,\gamma^5)\,+\, q^\mu\,(F_3^V-F_3^A\,\gamma^5)\,. \nonumber
\end{eqnarray}
We shall not write down rate expressions in terms of these form factors since
these have been worked out in great detail in Ref.~\cite{KKP}.

In the impulse approximation which is being employed in our approach, 
the matrix element $\Lambda^\mu$ is calculated according to
\begin{equation}\label{invampl}
\Lambda^\mu_{i \to f}(p,p')=
-12\,g_{i}g_{f}
\int\!\frac{d^4k_1}{(2\pi)^4i}\!\int\!\frac{d^4k_2}{(2\pi)^4i}\,
\phi_{i}(-k^2)\,\phi_{f}(-k^2)\,C_{i \to f}^\mu\,, 
\end{equation}
where
\begin{eqnarray*}
\hspace*{-.5cm}
C_{bc\to cc}^\mu\,&=&\, -\sqrt{2}\,\gamma^\alpha\gamma^5\,S_q(k_2)\,
\gamma^\beta\,S_c(k_1+k_2)\,\gamma^\alpha\, S_c(k_1+p') \, 
O^\mu\, S_b(k_1+p)\,\gamma^\beta\gamma^5 \,, \\
\hspace*{-.5cm}
C_{bc\to bs}^\mu\,&=&\,\gamma^\alpha\gamma^5\,S_b(k_1+p)\,
\gamma^\beta\gamma^5\,{\rm tr}\left(
S_s(k_2-q)\, O^\mu\, S_c(k_2)\,\gamma^\beta\, S_q(k_1+k_2)\,\gamma^\alpha 
\right)\,,\\
\hspace*{-.5cm}
C_{cc\to cs}^\mu\,&=&\,-\sqrt{2}\,\gamma^\beta\gamma^5\,S_c(k_1+p)\,
\gamma^\alpha\,S_c(k_2) \,O^\mu_+\,S_s(k_2+q)\gamma^\beta\, 
S_q(k_2-k_1)\,\gamma^\alpha\gamma^5 \,,\nonumber
\end{eqnarray*}
and where $O^\mu_+=\gamma^\mu(1+\gamma^5)$ and
$k^2\equiv k_1^2+(k_1+k_2)^2+k_2^2$. 
The quark propagator is chosen to have a local form
\begin{equation}\label{prop}
S_i(k)=\frac{1}{m_i-\not\! k} \hspace{1cm}
(i=u,d,s,c,b)
\end{equation}
with $m_i$ being a constituent quark mass. The vertex function $\phi_H$ is 
directly related to the Fourier-transform of the function $\Phi_H$ 
\begin{eqnarray*}
\tilde\Phi_H(p_1,...,p_4)&=&\int\!dx_1 ...\!\int\!dx_4\,
e^{i\sum\limits_{i=1}^4 x_i p_i}\Phi_H(x_1,...,x_4)\\
&=&(2\pi)^4\delta^{(4)}(\sum\limits_{i=1}^4 p_i)\,\phi_H(p_1,p_2,p_3).
\end{eqnarray*}
Generally, $\phi_H$ is a function of three momentum variables. 
However, in the impulse approximation employed in our approach, 
we assume that it only depends on the sum of relative momentum 
squared as indicated in Eq.~(\ref{invampl}).

The compositeness condition  reads
\begin{equation}\label{z=0}
Z_H=1-g^2_H\Sigma'(m_H)=0
\end{equation}
where $\Sigma'(m_H)$ is the derivative of baryon mass operator taken on its 
mass-shell. In the impulse approximation  Eq.~(\ref{z=0}) may be rewritten 
in a form suitable for the determination of the coupling constants:
\begin{eqnarray}
\hspace*{-.8cm}&&-12\, g^2_{q_1q_2q_3}
\int\!\frac{d^4k_1}{(2\pi)^4i}\!\int\!\frac{d^4k_2}{(2\pi)^4i}\,
\phi^2_{q_1q_2q_3}(-k^2)\,D^\mu_{q_1q_2q_3}\,|_{\,\not p=m_H}\,
=\,\gamma^\mu\,, \label{coupl}\\
\hspace*{-.8cm}&&D^\mu_{q_1q_2q_3}=\gamma^\alpha\gamma^5\,
S_{q_1}(k_1+p)\,\gamma^\mu\,S_{q_1}(k_1+p)\gamma^\beta\gamma^5\,
{\rm tr}\left(
S_{q_2}(k_1+k_2)\,\gamma^\alpha\, S_{q_3}(k_2)\,\gamma^\beta\right)\,,
\nonumber\\
\hspace*{-.8cm}&&(q_1q_2q_3)=(bcq)\,,(bsq)\,,(csq)\,,\nonumber\\
\hspace*{-.8cm}&&\nonumber\\
\hspace*{-.8cm}
&&D^\mu_{ccq}=\gamma^\alpha\gamma^5\, S_{q}(k_2)\,\gamma^\beta\gamma^5\,
{\rm tr}\left(\gamma^\alpha\,
S_{c}(k_1+p)\,\gamma^\mu\,S_{c}(k_1+p)\,\gamma^\beta\,
S_{c}(k_1+k_2)\right)\,. \nonumber
\end{eqnarray}
For the coupling constants we obtain $g_{bcq}=0.96$, $g_{bsq}=3.33$,
$g_{ccq}=2.63$, $g_{csq}=3.75$.

Next we turn to the calculation of the transition form factor. 
The calculational techniques are outlined in Ref.~\cite{finite}. 
The three main ingredients are
\begin{itemize}
\item  use of the Laplace transform of the vertex function 
$$
\Phi(z)=\int\limits_0^\infty\! ds\, \Phi_L(s)\, e^{-sz}
$$
\vspace{-0.2cm}
\item the  $\alpha$-transform of the denominator
$$
\frac{1}{m^2-(k+p)^2}=\int\limits_0^\infty\! d\alpha \,  
e^{-\alpha (m^2-(k+p)^2)}
$$
\vspace{-0.2cm}
\item  differential representation of the numerator
$$
\left(\,m+\not\! k +\not\! p¸\right)\, e^{kq}=
\left(\,m+\gamma^\mu \frac{\partial}{\partial q^\mu} +\not\! p\,\right)\, 
e^{kq}
$$
\end{itemize}
The calculation  of the transition form factors amounts to a two-loop
integrations. Four of the eight two-loop integrations are done analytically.
One ends up with 4-fold integrals which are not difficult to evaluate 
numerically. All calculations are done by using computer programs written 
in FORM for the manipulations of Dirac matrices and in FORTRAN for numerical 
evaluations. 

The common structure of the expressions for the form factors may be written as
\begin{equation}\label{expr}
\hspace*{-.5cm} v(q^2)=\int\limits_0^\infty\! dtt^3\! \int\! 
\frac{d^4\alpha}{|A|^2}\delta\left(1-\sum\limits_{i=1}^4\alpha_i\right)\
\left(F(z) W_0-\frac{1}{2} F_1(z)W_1+\frac{1}{4} F_2(z)W_2\right)
\end{equation}
where
\begin{eqnarray*}
F(z)&=&\phi_{\rm in}(z)\phi_{\rm out}(z)\,, \hspace{0.5cm}
F_i(z)=\int\limits_0^\infty d\tau \tau^i F(z+\tau)\,,\\
z&=&t\left(\sum\limits_{i=1}^4\alpha_i m^2_{q_i}-\alpha_1 p^2-\alpha_2 
p'^2\right)+t^2 A_{11}^{-1} (\alpha_1 p-\alpha_2 p')^2\,.
\end{eqnarray*}
and where the matrix $A$ is defined as

\[
A=\left(
\begin{array}{cc}
\mbox{$2+t(\alpha_1+\alpha_2+\alpha_3)$}
&\mbox{$1+t\alpha_3$}\\
\mbox{$1+t\alpha_3$}&\mbox{$2+t(\alpha_3+\alpha_4)$}
\end{array}
\right) 
\]

The functions $W_i$ contain  integration variables and masses.
One has to emphasize that the above  expressions are valid for any
vertex functions decreasing rapidly enough in the Euclidean region. 
Since the quark masses satisfy the confinement constraint 
$m_H<\sum\limits_{i=1}^3 m_{q_i}$ all form factors are real.

Before presenting our numerical results we need to specify our values
for the constituent quark masses and shapes of the vertex functions.
As concerns the vertex functions, we found a good description of various
physical quantities \cite{Bc,model,RTQM,finite} adopting a Gaussian form.
Here we apply the same procedure using 
$
\phi_H(k^2_E)=\exp\{-k^2_E/\Lambda_H^2\}
$
in the Euclidean region. The magnitude of $\Lambda_H$ characterizes the
size of the vertex function and is an adjustable parameter in our model.
The $\Lambda_H$ parameters in the meson sector were determined \cite{model}
by a least-squares fit to experimental data and lattice determinations.
The nucleon $\Lambda_N$ parameter was determined from the best description of 
the electromagnetic properties of the nucleon \cite{model}. 
The $\Lambda_H$ parameters for baryons with one heavy quark (bottom or charm)
were determined by analyzing available experimental data on bottom and charm 
baryon  decays. Since there is no  experimental information on the properties 
of double heavy baryons yet we use the simple observation that the magnitude 
of $\Lambda_H$ is increasing  with the mass value of the hadron whose shape 
it determines. Keeping in mind that $\Lambda_N=1.25$ GeV, 
$\Lambda_{Qqq}=1.8$ GeV and $\Lambda_{B_c}=2.43$ GeV, we simply choose the 
value of $\Lambda_{QQq}=2.5$ GeV for the time being. We found that variations 
of this value by 10 $\%$  does not much affect the values of form factors. 
We employ the same values for the quark masses (see, Eq.(\ref{quarks})) 
as have been used previously for the description of light and heavy 
baryons \cite{model,RTQM}. We thus use
\begin{equation}\label{quarks}
\begin{array}{ccccc}
m_u & m_s & m_c & m_b & \\
\hline
$\ \ 0.420\ \ $ & $\ \ 0.570\ \ $ &  $\ \ 1.67\ \ $ & $\ \ 5.06\ \ $  &
\,\,{\rm (GeV)}\\
\end{array}
\end{equation}
The resulting form factors are {\it approximated} by the interpolating form
\begin{equation}\label{approx}
f(q^2)=\frac{f(0)}{1-a_1 q^2+a_2 q^4}
\end{equation}
It is interesting that for most of the form factors the numerical fit  
values of $a_1$ and $a_2$ obtained from the interpolating form~(\ref{approx}) 
are such that the form factors can be represented by dipole formula
\begin{equation}\label{dipole}
f(q^2)\approx d(q^2)=\frac{f(0)}{(1-q^2/m^2_V)^2}
\end{equation}
The values of $m_V$ in the dipole representation are very close to the values 
of the appropriate lower-lying $(\bar q q')$ vector mesons 
($m_{D^\ast_s}$=2.11 GeV for (c-s)-transitions and 
$m_{B_c^\ast}\approx m_{B_c}$=6.4 GeV for (b-c)-transitions). In Fig.1 
we show two representative form factors and their dipole approximations. 
We have shown in Ref.~\cite{Bc} that  the form factors of the CKM-enhanced 
semileptonic $B_c$-decays  may be approximated by a monopole function. 
It is gratifying to see  that our relativistic quark model with the Gaussian 
vertex function and free quark propagators reproduces the monopole in the 
meson case and the dipole in the baryon case for most of the form factors. 

Finally, in Table 2 we present our predictions for the decay rates and 
compare them with the free quark decay width which is the leading 
contribution to the semileptonic inclusive width $(x=m^2_{Q_f}/m^2_{Q_i})$
\begin{equation}
\Gamma_{\rm 0}(i\to f)=|V_{if}|^2\,\frac{G_F^2\,m_{Q_i}^5}{192\,\pi^3}\,
\left(1-8\,x+8\,x^3-x^4-12\,x^2\,\ln x\right)\,.
\end{equation}
The nonleading corrections to the leading order rate Eq.(12) lower the 
inclusive rate by approximately 1\% and 15\% in the $b \rightarrow c$ 
and $c \rightarrow s$ case, respectively (see e.g. \cite{KM}).  
In the numerical evaluation of the inclusive rate Eq.(12) we used
current pole mass values $m_b=4.8$~GeV and $m_c=1.325$~GeV \cite{Pivo}.
There are many values quoted in the literature for the current pole mass
of the strange quark. For the sake of definiteness we take $m_s=0.15$~GeV. 
Note that in both $cc\to cs$ and $bc\to cc$ decays there is
an additional factor of 2 due to the fact that there are two c-quarks
in the double charmed baryon in the initial and final state,  
respectively \cite{Likh,GMS}.
One notes that the rates for the exclusive modes $\Xi_{bc}\to\Xi_{cc}+l\bar\nu$
and $\Xi_{bc}\to\Xi_{bs}+l\bar\nu$ are rather small when compared
to the total semileptonic inclusive rate estimated by Eq.(12). The remaining
part of the inclusive rate would have to be filled in by decays into excited
or multi-body baryonic states. Note that the smallness of the exclusive/inclusive
ratio of the above exclusive modes
markedly differs from that of the mesonic semileptonic
$b\to c$ transitions, where the exclusive transitions to the ground state 
S-wave mesons $B\to D,D^\ast$ make up approximately 66$\%$ of the total 
semileptonic $B\to X_c$ rate \cite{PDG}. For $\Lambda_b\to\Lambda_c$ 
transitions one expects even higher semileptonic exclusive-inclusive ratio 
of amount 80$\%$ \cite{KM}. Note that the rate for $\Xi_{bc}\to\Xi_{cc}+l\bar\nu$ is of 
the same order of magnitude as the rates calculated for the corresponding 
double heavy mesonic decays $B_c\to\eta_c+l\bar\nu$ and 
$B_c\to J/\Psi+l\bar\nu$ \cite{Bc}. The QCD sum rule and potential model 
predictions for the rates of $\Xi_{bc}\to\Xi_{cc}+l\bar\nu$ and 
$\Xi_{bc}\to\Xi_{cs}+l\bar\nu$ given in \cite{Likh} exceed our rate 
predictions by factors of 10 and 3 respectively. In fact, the exclusive 
semileptonic rates given in \cite{Likh} tend to saturate the inclusive 
semileptonic rates calculated from Eq.(12) as given in Table 2. 

In Table 3 we present  values for the invariant form factors at 
$q^2_{\rm min}=0$ and $q^2_{\rm max}=(m_i-m_f)^2$. Note that the values of 
the axial vector form factor $F_1^A$ are rather small for the two decays 
$\Xi_{bc}\to\Xi_{cc}+l\bar\nu$ and $\Xi_{bc}\to\Xi_{bs}+l\bar\nu$. 
This provides for a partial explanation of why the rates of these two modes 
are small compared to the inclusive semileptonic rate. Also the zero recoil 
values of the vector form factors are significantly below the value of one 
which one would expect from a naive application of the heavy quark limit. 
The smallness of the vector form factors provide for the remaining 
explanation of the smallness of the predicted respective rates. We mention 
that the QCD sum rule and potential model estimates of the zero recoil values 
of both the vector and axial vector form factors $F_1^V$ and $F_1^A$ given
in \cite{Likh} are close to one. In the model of \cite{Likh} the form factors 
$F_2^V$ and $F_2^A$ are set to zero. In our approach we find that the 
numerical values of $F_2^V$ and $F_2^A$ are quite small compared to those 
of $F_1^V$ and $F_1^A$ in all cases when expressed in terms of the mass 
scale $(m_i+m_f)$.

It is interesting to compare our full zero recoil results with those of 
a naive spectator quark model calculation where the zero recoil values of 
the form factors result from a simple overlap calculation. For the spin-flavor 
wave functions we use the $SU(12)$ spin-flavor wave functions of the naive 
spectator quark model (6 flavors $\times$ 2 spin projections) as extended from
its original  $SU(6)$ version to include the heavy flavors. As mentioned 
before this does not imply that one is assuming $SU(12)$ symmetry to hold for 
the transitions since one is using physical quark and baryon masses which 
badly break the $SU(12)$ symmetry. Instead one uses $SU(12)$ symmetry only 
to construct the spin-flavor wave functions of the heavy and double heavy 
baryons. In the $SU(12)$ naive spectator quark model the spin-flavor wave 
functions $|B>$ of  $\Xi_{bcq}$, $\Xi_{bsq}$ and $\Xi_{ccq}$ baryons with 
diquarks in the symmetric state and positive $(+ 1/2)$ spin projection are 
given by  
\begin{eqnarray}
\hspace*{-.75cm} 
|\Xi_{bcq}; \uparrow> &=& \frac{1}{6\sqrt{2}} 
|2qcb + 2cqb - qbc - cbq - bqc - bcq> 
|\uparrow \downarrow \uparrow + \downarrow \uparrow \uparrow - 
2 \uparrow \uparrow \downarrow > \nonumber\\
\hspace*{-.75cm} 
|\Xi_{bsq}; \uparrow> &=& \frac{1}{6\sqrt{2}} 
|2qsb + 2sqb - qbs - sbq - bqs - bsq> 
|\uparrow \downarrow \uparrow + \downarrow \uparrow \uparrow - 
2 \uparrow \uparrow \downarrow > \nonumber\\ 
\hspace*{-.75cm}  
|\Xi_{ccq}; \uparrow> &=& \frac{1}{2\sqrt{2}} 
|qcc + cqc - 2ccq> 
|\uparrow \downarrow \uparrow + \downarrow \uparrow \uparrow - 
2 \uparrow \uparrow \downarrow > \nonumber 
\end{eqnarray}
The values of $F_1^V$ and $F_1^A$ form factors at zero recoil can be 
calculated from the matrix elements  
\begin{eqnarray}
F_1^V = <B^\prime|\sum\limits_{i=1}^3 [I_{fl}]^{(i)} |B>
\hspace*{.3cm} \mbox{and} \hspace*{.3cm}
F_1^A = <B^\prime|\sum\limits_{i=1}^3 [\sigma_3 I_{fl}]^{(i)} |B> 
\nonumber
\end{eqnarray}
where $\sigma_3$ is the Pauli (third component) spin matrix and $I_{fl}$ 
is the flavor matrix responsible for the s.l. transitions. 

In the Table 4 we present the results for the $F_1^V$ and $F_1^A$ form 
factors at zero recoil calculated in the naive spectator quark model. 
It is evident that the spectator quark model prediction of a vanishing axial 
vector coupling $F_1^A(0)=0$ for the decay $\Xi_{bc}\to\Xi_{cc}+l\bar\nu$ is 
close to the suppressed value of $F_1^A(0)=-0.091$ in the full calculation. 
The ratios of the axial and vector couplings in the decay 
$\Xi_{bc}\to\Xi_{bs}+l\bar\nu$ are $F_1^A(0)/F_1^V(0)=1/2$ (naive quark model) 
and $F_1^A(0)/F_1^V(0) \approx 1/6$ (our approach). 
For $\Xi_{cc}\to\Xi_{cs}+l\bar\nu$ decay one has $F_1^A(0)/F_1^V(0)=1$ 
(naive spectator quark model) and $F_1^A(0)/F_1^V(0) \approx 1.3$ 
(our approach). The different normalizations of axial and vector constants 
obtained in our approach compared to the naive spectator quark model result 
and the suppression of the ratio $F_1^A(0)/F_1^V(0)$ in the decay 
$\Xi_{bc}\to\Xi_{bs}+l\bar\nu$ in the full calculation can be explained by 
relativistic effects and nontrivial heavy quark/baryon mass dependence of 
the relevant matrix elements. 

In order to test the sensitivity of our results on  various choices of the 
bottom and charm quark masses, we have varied their values within 
a reasonable range. From the confinement constraints one obtains  
lower permissible values for heavy quark masses: 
$m_c\ge (m_{\Xi_{cc}}-m_u)/2= 1.60$ GeV and
$m_b\ge (m_{\Xi_{bc}}-m_c-m_u)=4.98$ GeV.
The upper values were found from the experimental bounds for the 
$\Lambda_b\to \Lambda_c e^-\bar \nu$ and $\Lambda_c\to \Lambda_s e^+ \nu$ 
decay rates: $m_c\le 1.72$ GeV and $m_b\le 5.25$ GeV \cite{finite}. 
We have calculated  the values of  decay rates of double baryons for three 
set of quark masses in Table 5. The decay rates do not change significantly 
in the chosen regions of the heavy quark masses.
\vspace*{1cm}

The visit of M.A.I. to Mainz University was supported by the DFG (Germany).
This work was supported in part by the Heisenberg-Landau Program,
the Russian Fund of Basic Research 01-02-17200 and the DFG (FA67/25-1).

\newpage 

\begin{table}[t2]
\caption{Calculated decay widths of lowest lying $J^P=1/2^+$ double heavy 
$\Xi$-baryons. Inclusive widths are calculated using the current
quark pole masses.}
\vspace*{0.2cm}
\def\arraystretch{1.85}
\begin{center}
\begin{tabular}{|c|c|r|}
\hline
    & \multicolumn{2}{|c|} {Decay widths, ps$^{-1}$} \\
\cline{2-3} Mode  & RTQM & Inclusive width  \\
\hline\hline
$\Xi_{\,\rm bc}\,\to\,\Xi_{\,\rm cc}\,+\, l\,\bar\nu$
& \,0.012\, & 
 $2\cdot\Gamma_0(b\to c)$\,=\,0.162 \\
\hline
$\Xi_{\,\rm bc}\,\to\, \Xi_{\,\rm bs}\,+\, l\,\bar\nu$
& \,0.043 \, & 
$\Gamma_0(c\to s)$\,=\, 0.122\\
\hline
$\Xi_{\,\rm cc}\,\to\, \Xi_{\,\rm cs}\,+\, l\,\bar\nu$
& \,0.224\,  &
$2\cdot\Gamma_0(c\to s)$\,=\,0.244\,\\
\hline 
\end{tabular}
\end{center}
\end{table}

\begin{table}[t3]
\caption{Values of $F_1^V$ and $F_1^A$ form factors at maximum and zero 
recoil.}
\vspace*{0.2cm}
\def\arraystretch{1.85}
\begin{center}
\begin{tabular}{|l|cc|cc|cc|}
\hline
     & \multicolumn{2}{|c|} {$\Xi_{bc}\to \Xi_{cc}$} & 
       \multicolumn{2}{|c|} {$\Xi_{bc}\to \Xi_{bs}$} & 
       \multicolumn{2}{|c|} {$\Xi_{cc}\to\Xi_{cs}$}  \\
\cline{2-7} & $F^V_1$ & $F^A_1$ & $F^V_1$ & $F^A_1$  & $F^V_1$ & $F^A_1$  \\
\hline\hline
$q^2=0$ &\, 0.46\,  &\, -0.091\, &\, 0.39\, &
\, 0.061 \,& \,0.47\, &\, 0.61\, \\
\hline
$q^2=q^2_{\rm max}$ &\, 0.83\, &\, -0.086\, &\, 0.58\, &\, 0.065 \,& 
\, 0.59\, &\, 0.77\, \\ 
\hline 
\end{tabular}
\end{center}
\end{table}

\begin{table}[t4]
\caption{Values of $F_1^V$ and $F_2^V$ form factors at zero 
recoil in the naive spectator quark model.}
\vspace*{0.2cm}
\def\arraystretch{1.85}
\begin{center}
\begin{tabular}{|c|c|c|c|}
\hline
Quantity & $\Xi_{bcq}\to\Xi_{ccq}$ & $\Xi_{bcq}\to\Xi_{bsq}$ 
         & $\Xi_{ccq}\to\Xi_{csq}$ \\
\hline\hline 
$F_1^V$  & $1/\sqrt{2}$ & $1$   & $1/\sqrt{2}$ \\
\hline
$F_1^A$  & $0$          & $1/2$ & $1/\sqrt{2}$ \\
\hline
\end{tabular}
\end{center}
\end{table}

\begin{table}[t5]
\caption{Calculated decay widths of double heavy $\Xi$-baryons
for three sets of quark masses.}
\vspace*{0.2cm}
\def\arraystretch{1.9}
\begin{center}
\begin{tabular}{|c|c|c|c|}
\hline
Decay & $m_c=1.60$ GeV & $m_c=1.67$ GeV & $m_c=1.72$ GeV \\
      & $m_b=4.98$ GeV & $m_b=5.06$ GeV & $m_b=5.25$ GeV  \\
\hline\hline
$\Gamma\left(\Xi_{\,\rm bc}\,\to\,\Xi_{\,\rm cc} + l\,\bar\nu\right)$,
${\rm ps^{-1}}$   & 0.0102 & 0.0117 & 0.0123 \\
\hline
$\Gamma\left(\Xi_{\,\rm bc}\,\to\,\Xi_{\,\rm bs} + l\,\bar\nu\right)$,
${\rm ps^{-1}}$   & 0.0498 & 0.0432 & 0.0371 \\
\hline
$\Gamma\left(\Xi_{\,\rm cc}\,\to\,\Xi_{\,\rm cs} + l\,\bar\nu\right)$,
${\rm ps^{-1}}$   & 0.258 & 0.224 & 0.208 \\
\hline
\end{tabular}
\end{center}
\end{table}

\newpage

\begin{figure}[t1]
\begin{center}
\begin{tabular}{c}
\epsfig{figure=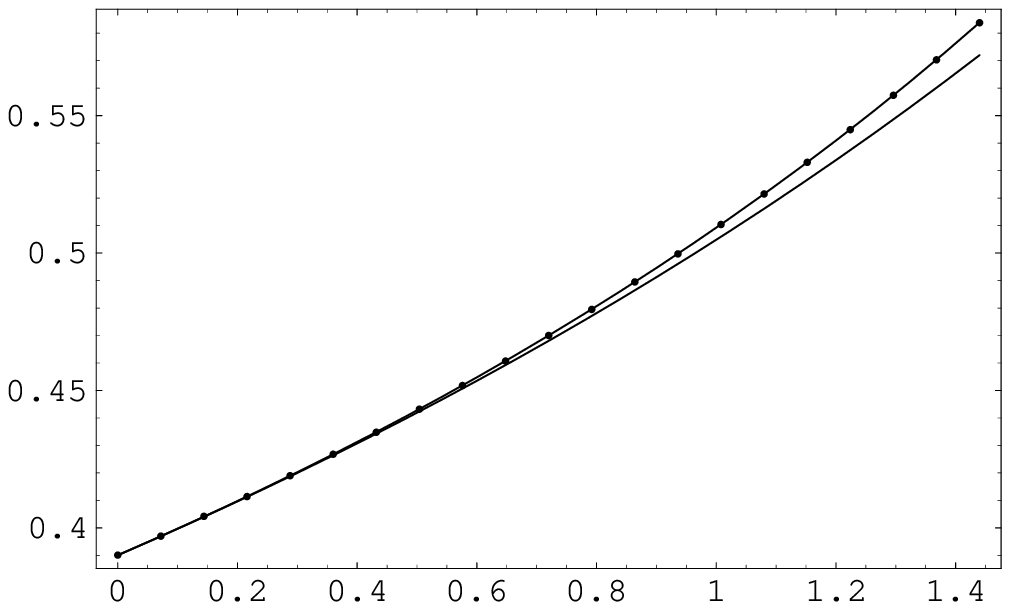,height=8.3cm}\\
\epsfig{figure=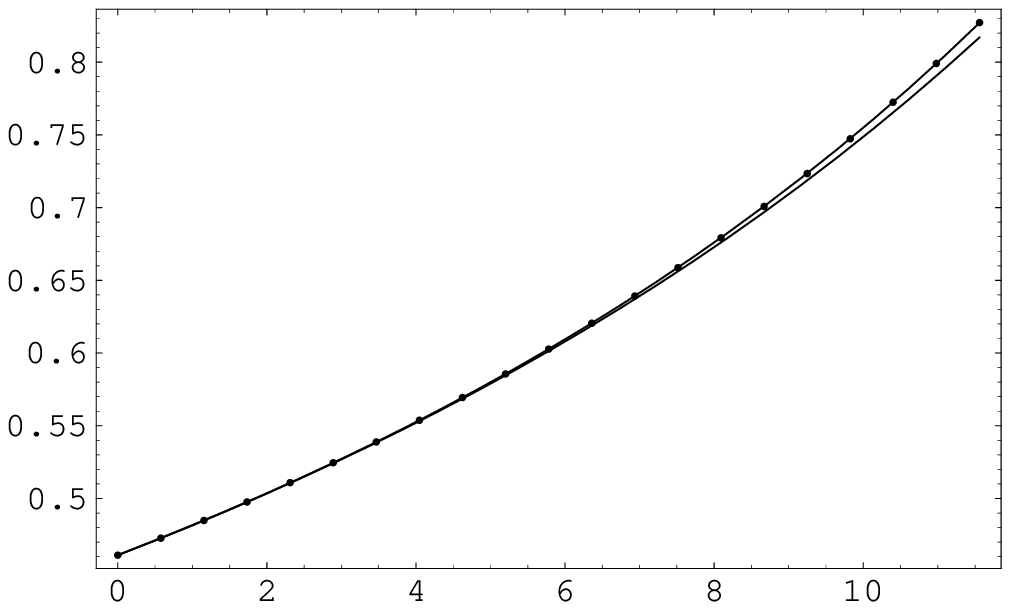,height=8.3cm}
\end{tabular}
\end{center}
\caption{
Upper panel:  Form factor $F_1^V (q^2)$ (solid-dotted line)
for $bc\to bs$ transitions and its dipole 
approximation (solid line) with $m_{cs}=2.88$ GeV;
Lower panel:  Form factor $F_1^V (q^2)$ (solid-dotted line)
for  $bc\to cc$ transitions and its dipole 
approximation (solid line) with  $m_{bc}=6.81$ GeV.
}
\end{figure}

\vspace*{1cm}
\newpage 

\begin{thebibliography}{999}
%
%
\bibitem{CDF} F. Abe {\it et al.} CDF Collaboration, 
Phys. Rev. Lett. {\bf 81} (1998) 2432;
Phys. Rev. D {\bf 58} (1998) 112004. 
%
%
\bibitem{Likh} V.V. Kiselev and A.K. Likhoded,
``Baryons with two heavy quarks'', 
hep-ph/0103169.
\bibitem{KKP} J.G. K\"orner, M. Kr\"amer and D. Pirjol,
Prog. Part. Nucl. Phys. {\bf 33} (1994) 787.
\bibitem{GMS} B. Guberina, B. Melic and H. Stefancic,
Eur. Phys. J. C {\bf 9} (1999) 213.
%
%
\bibitem{Bc} M.A. Ivanov, J.G. K\"orner and P. Santorelli, 
Phys. Rev. D {\bf 63} (2001) 074010. 
%
%
\bibitem{model} M.A. Ivanov, M.P. Locher and V.E. Lyubovitskij,
Few-Body Syst. {\bf 21} (1996) 131;
M.A. Ivanov and V.E. Lyubovitskij, Phys. Lett. B {\bf 408} (1997) 435; 
M.A. Ivanov and P. Santorelli, Phys. Lett. B {\bf 456} (1999) 248. 
%
%
\bibitem{SWH} A. Salam, Nuovo Cim. {\bf 25} (1962) 224;
S. Weinberg, Phys. Rev. {\bf 130} (1963) 776;
K. Hayashi et al., Fort. Phys. {\bf 15} (1967) 625.
\bibitem{EI} G.V. Efimov and M.A. Ivanov, 
Int. J. Mod. Phys. A {\bf 4} (1989) 2031; 
{\it ``The Quark Confinement Model of Hadrons''}, IOP, London, 1993.
%
%
\bibitem{DSE} C.D. Roberts and A.G. Williams, Prog. Part. Nucl. Phys.
{\bf 33} (1994) 477. 
\bibitem{DSEH} M.A. Ivanov, Yu.L. Kalinovsky and C.D. Roberts,
Phys. Rev. D {\bf 60} (1999) 034018.
%
%
\bibitem{RTQM} M.A. Ivanov, V.E. Lyubovitskij, J.G. K\"{o}rner and P. Kroll,
Phys. Rev. D {\bf 56} (1997) 348;
M.A. Ivanov, J.G. K\"{o}rner, V.E. Lyubovitskij and A.G. Rusetsky, 
Phys. Rev. D {\bf 57} (1998) 5632; D {\bf 60} (1999) 094002.
%
%
%
\bibitem{Gatto} A. Deandrea {\it et al.} Phys. Rev. D {\bf 58} (1998) 034004.
\bibitem{finite} M.A. Ivanov, J.G. K\"{o}rner, V.E. Lyubovitskij and 
A.G. Rusetsky, Phys. Lett. B {\bf 476} (2000) 58. 
%
%
\bibitem{Faust}D. Ebert {\it et al.} Z. Phys. C {\bf 76} (1997) 111.
\bibitem{KM}J.G. K\"{o}rner and B. Melic, Phys. Rev. D {\bf 62} (2000) 074008. 
\bibitem{Pivo}A.A. Penin and A.A. Pivovarov, 
Nucl. Phys. B {\bf 549} (1999) 217.
\bibitem{PDG} Particle Data Group, D.E. Groom {\it et al.} Eur. Phys. J.
C {\bf 15} (2000) 1. 
\end{thebibliography}
\end{document}